\documentclass[journal]{IEEEtran}
\ifCLASSINFOpdf

\else

\fi

\usepackage{amsmath}
\usepackage{amsfonts}
\usepackage{hyperref}
\usepackage{cleveref}
\usepackage{cite}
\usepackage{xcolor}
\usepackage{graphicx}

\DeclareMathOperator*{\argmin}{arg\,min}

\begin{document}

\title{Feasibility of iMagLS-BSM - ILD Informed Binaural Signal Matching with Arbitrary Microphone Arrays}

\author{Or~Berebi,~\IEEEmembership{Student Member,~IEEE,}
        Zamir~Ben-Hur,~\IEEEmembership{Member,~IEEE,}
        David~Lou~Alon,~\IEEEmembership{Member,~IEEE,}
        and Boaz~Rafaely,~\IEEEmembership{Senior Member,~IEEE}% <-this % stops a space
\thanks{O. Berebi and B. Rafaely are with the School of Electrical and Computer Engineering, Ben-Gurion University of the Negev, Beer-Sheva 84105, Israel.}% <-this % stops a space
\thanks{Z. Ben-Hur and D. Lou~Alon are with Reality Labs Research, Meta, 1 Hacker Way, Menlo Park, CA 94025, USA.}}% <-this % stops a space

\maketitle

\begin{abstract}
Binaural reproduction for headphone-centric listening has become a focal point in ongoing research, particularly within the realm of advancing technologies such as augmented and virtual reality (AR and VR). The demand for high-quality spatial audio in these applications is essential to uphold a seamless sense of immersion. However, challenges arise from wearable recording devices equipped with only a limited number of microphones and irregular microphone placements due to design constraints. These factors contribute to limited reproduction quality compared to reference signals captured by high-order microphone arrays. This paper introduces a novel optimization loss tailored for a beamforming-based, signal-independent binaural reproduction scheme. This method, named iMagLS-BSM incorporates an interaural level difference (ILD) error term into the previously proposed binaural signal matching (BSM) magnitude least squares (MagLS) rendering loss for lateral plane angles. The method leverages nonlinear programming to minimize the introduced loss. Preliminary results show a substantial reduction in ILD error, while maintaining a binaural magnitude error comparable to that achieved with a MagLS BSM solution. These findings hold promise for enhancing the overall spatial quality of resultant binaural signals.
\end{abstract}


\IEEEpeerreviewmaketitle

\section{Introduction}
Binaural audio reproduction has become a very active research field in recent years, with an increasing number of applications in particular for virtual and augmented reality~\cite{rafaely2022spatial}. Binaural audio is well suited to these applications since it enables the listener to experience three-dimensional soundscapes reproduced by headphones. A popular approach to capturing an acoustic scene and reproducing it in binaural audio is the popular Ambisonics format~\cite{zotter2019ambisonics}. In Ambisonics, the acoustic scene is typically captured by a spherical microphone array; the sound field representation is then filtered by the listener head-related transfer function (HRTF)~\cite{xie2013head}, forming the binaural signals. However, the need for a spherical microphone array limits its applicability in scenarios where more flexibility in microphone placement is needed, such as wearable microphone arrays or microphone arrays mounted on electronic consumer products.

One approach to binaural reproduction from arbitrary arrays is the binaural signals matching (BSM) approach~\cite{madmoni2020beamforming,rasumow2011robustness,mccormack2023six}. This usually refers to the estimation of binaural signals while minimizing the mean-squared error (MSE) of magnitude least-squares error (MagLS) by matching the array steering vectors to the HRTFs using a linear formulation. This approach showed good promise. However, performance is still lacking when comparing its resulting binaural signals to high quality signals, in terms of perceived spatial quality, especially for arrays where microphones are located far from the listener's ears~\cite{mccormack2023six}. An alternative approach for binaural reproduction with arbitrary microphone arrays could be the family of parametric approaches~\cite{fernandez2022enhancing, mccormack2022parametric, mccormack2019parametric}. These approaches are signal-dependent, meaning that performance depends on the estimation accuracy of scene parameters such as the direction-of-arrival, diffuseness of the sound field, and the sparsity assumption of the sound sources in the time-frequency domain. This work will focus on signal-independent methods due to this reason.

In this paper, a novel BSM-based reproduction approach is presented. Drawing inspiration from a previous method presented in~\cite{berebi2023imagls}, which optimized low-order spherical harmonics (SH) coefficients of an HRTF for both magnitude and interaural level difference (ILD), our approach extends beyond spherical arrays. The suggested method aims to address the perceived quality gap of the previously suggested BSM method by introducing an ILD error term to the previously suggested mean square error (MSE) and magnitude error terms. Objective evaluation, employing simulated microphone arrays and recorded HRTFs, validates the efficacy of the proposed approach, demonstrating a noticeable reduction in ILD error alongside a marginal increase in magnitude error.

\section{Binaural Signal Matching}

This section provides the necessary mathematical background for understanding the problem of signal-independent binaural signal matching from arbitrary microphone arrays, assuming a narrow-band model.

Consider a sound-field composed of $Q$ far-field sources arriving from directions $\{\Omega_q = (\theta_q, \phi_q)\}_{q=1}^{Q}$. Binaural signals can be formulated for this case as
\begin{equation}\label{traget_signals}
    p^{l,r}(f) = [\mathbf{h}^{l,r}(f)]^T\mathbf{s}(f),
\end{equation}
where $\mathbf{s}(f)=[s_1(f, \Omega_1), \dots, s_Q(f, \Omega_Q)]^T$ represents the $Q$-length source signal vector associated with the corresponding source direction of arrival (DOA) $\Omega_q$. The vector $\mathbf{h}^{l,r}(f)=[h^{l,r}_1(f, \Omega_1), \dots, h^{l,r}_Q(f, \Omega_Q)]^T$ represents the left and right ear head-related transfer function (HRTF), with each entry being the frequency response for the far-field source at direction $\Omega_q$.

Binaural signal matching (BSM) refers to a beamforming-based approach for rendering binaural signals, where the microphone signals are directly matched to the left and right binaural signals~\cite{madmoni2020beamforming}
\begin{equation}\label{est_signals}
    z^{l,r}(f) = [\mathbf{c}^{l,r}(f)]^H\mathbf{x}(f),
\end{equation}
where $\mathbf{c}^{l,r}(f)$ depicts the $M$-length BSM filter coefficients vector for the left and right ear. The signal model for the microphone array signal is given by
\begin{equation}\label{mic_model}
    \mathbf{x}(f) = \mathbf{V}(f)\mathbf{s}(f) + \mathbf{n}(f) 
\end{equation}
Here, $\mathbf{x}(f) = [ \, x_1(f), \dots,x_M(f) ] ^T$ denotes a vector of length $M$ representing the microphone signals, with $(.)^T$ denoting the transpose operation. The matrix $\mathbf{V}(f) = [ \, \mathbf{v}_1(f),\dots, \mathbf{v}_Q(f) ] $ is an $M \times Q$ matrix, where the $q$-th column contains the steering vector for the $q$-th source. The $M$ length vector $\mathbf{n}(f)$ represents noise at the microphones.

The BSM coefficients $\mathbf{c}^{l,r}(f)$ are calculated based on a minimization criterion. Two criteria are considered as baselines in this study. The first is the mean square error (MSE) minimization, for which the coefficients are computed using
\begin{equation}\label{problem:mse}
    \mathbf{c}^{l,r}_{mse} =  \argmin_{\mathbf{c}^{l,r}} \mathbb{E} \left[ | p^{l,r}(f) - z^{l,r}(f) |^2 \right].
\end{equation}
Since binaural signals are estimated with limited prior knowledge, the sound sources are assumed to be spatially white, or diffuse, while the noise is assumed uncorrelated and identically distributed between microphones. This assumption enables deriving an analytic solution to Eq.~\ref{problem:mse}, as shown in~\cite{madmoni2020beamforming}
\begin{equation}\label{solution:mse}
    \mathbf{c}^{l,r}_{mse} = (\mathbf{V}\mathbf{V}^H + \frac{\sigma_n^2}{\sigma_s^2}\mathbf{I}_M)^{-1} \mathbf{V} [\mathbf{h}^{l,r}]^*,
\end{equation}
where the source and noise variances are denoted by $\sigma_s^2$ and $\sigma_n^2$ respectively. The matrix $\mathbf{V}(f) = [ \, \mathbf{v}_1(f),\dots, \mathbf{v}_K(f) ] $ is an $M \times K$ matrix, where the $k$-th column contains the steering vector for the $k$-th source. Due to a lack of knowledge regarding the number and directions of the sources in the sound-field, the MSE solution is signal-independent, and a dense grid of $K$ source directions is used to derive $\mathbf{c}^{l,r}_{mse}$; both $\mathbf{V}$ and $\mathbf{h}^{l,r}$ are assumed to be known in these $K$ directions. The MSE solution was evaluated in~\cite{madmoni2020beamforming} on a small semi-circular microphone array and was shown to result in a relatively low normalized MSE only up to $2$kHz, suggesting relatively poor performance for the entire audible range. Additionally, these results were verified in~\cite{mccormack2023six}, where the MSE solution for the BSM was evaluated in a listening test employing an eyeglasses-mounted seven-microphone array. The MSE solution was shown to be outperformed in terms of audible perception by the following method.

The second approach is grounded in perceptually motivated magnitude least squares (MagLS) optimization, as discussed in works such as~\cite{zotter2019ambisonics, deppisch2021end, strutt1907our}. In this method, the focus lies on matching the estimated signals solely based on their magnitudes at higher frequencies to the target signal. This contrasts with the strategy employed in Eq.~\ref{problem:mse}, where both magnitude and phase are considered. Mathematically, this is expressed as:
\begin{equation}\label{problem:mls}
    \mathbf{c}^{l,r}_{mls} =  \argmin_{\mathbf{c}^{l,r}} \mathbb{E} \left[ |\,\, |p^{l,r}(f)| - |z^{l,r}(f)| \,\, |^2\right].
\end{equation}
Regrettably, this minimization of magnitude least squares poses a non-convex problem, lacking a closed-form solution. However, diverse approaches for identifying a local minimum to Eq.~\ref{problem:mls} are detailed in~\cite{kassakian2006convex}, accompanied by a theoretical analysis of their efficacy. The study in~\cite{mccormack2023six} demonstrates that the MagLS solution surpasses the MSE solution in terms of audio perception. Nevertheless, it does not establish perceptual indistinguishability from the reference signal, suggesting room for potential enhancement.
\begin{figure}
 \centerline{\framebox{
 \includegraphics[width=7.8cm]{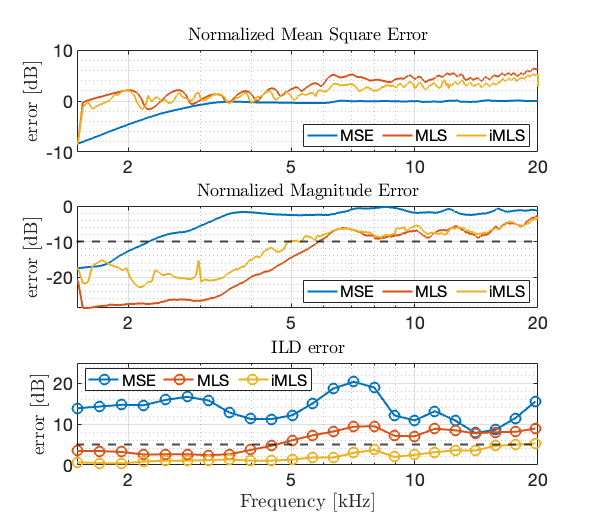}}}
 \caption{Complex error (top) Magnitude error (middle) and ILD error (bottom) as a function of frequency, averaged over the directions.}
 \label{fig:eMAG_eILD_freq}
\end{figure}

\section{Proposed Method}

We introduce the ILD-informed magnitude least squares optimization (iMagLS-BSM) to address the persisting performance gap in signal-independent binaural rendering compared to the state-of-the-art MagLS-BSM method. Inspired by the approach proposed in~\cite{berebi2023imagls}, our method directly optimizes the Interaural Level Difference (ILD) error alongside magnitude minimization. This is motivated by the potential achievement of low ILD error even in frequency ranges where high magnitude error persists. With this respect it should be noted that low magnitude error achieved for both ears, lead directly to a low ILD error and so in this case, the addition of and ILD error is redundant.

The novel optimization scheme is expressed as
\begin{equation}\label{C_imls}
    \mathbf{C}_{imls} =  \argmin_{\mathbf{C}} \mathbb{E} \left[\frac{1}{2}\left(\epsilon^{l}_{mls}+\epsilon^{r}_{mls}\right) + \lambda \epsilon_{ILD} \right],
\end{equation}
where $\mathbf{C}=[\mathbf{c}^l(f),\mathbf{c}^r(f)]$ represents a $M \times 2$ BSM filter coefficients matrix, with each column containing the coefficients for the left and right ears. The magnitude and ILD errors are weighted using the regularization term $\lambda \in \mathbb{R}$.

The left and right magnitude error is defined as 
\begin{equation}\label{eq:emag}
    \epsilon^{l,r}_{mls}(f) =  \mathbb{E} \left[ |\,\, |p^{l,r}(f,\Omega_k)| - |z^{l,r}(f,\Omega_k)| \,\, |^2 \right],
\end{equation}
depicting the magnitude error between the target binaural signal $p^{l,r}(f,\Omega_k)$ (defined in Eq.~\ref{traget_signals}) and the BSM estimation $z^{l,r}(f,\Omega_k)$ (defined in Eq.~\ref{est_signals}), averaged over $K$ directions. Both signals are evaluated for a single far-field sound source arriving from direction $\{\Omega_k = (\theta_k, \phi_k)\}_{k=1}^{K}$, representing a dense spherical grid of $K$ directions.

The direction-specific ILD error term $\epsilon_{ILD}$ is defined as 
\begin{equation}\label{eq:eILD}
    \epsilon_{ILD}(f_0,\Omega_l) = | ILD_{tgt}(f_0,\Omega_l) - ILD_{est}(f_0,\Omega_l)|,
\end{equation}
where the frequency-averaged ILD of the target binaural signals is defined as~\cite{xie2013head}
\begin{equation}\label{ILD:target}
    \text{ILD}_{tgt}(f_0,\Omega_l) = 10 \log_{10}\frac{\int_{f_1}^{f_2} G(f_0,f)|p^l(\Omega_l,f)|^2 df}{\int_{f_1}^{f_2} G(f_0,f)|p^r(\Omega_l,f)|^2 df},
\end{equation}
and the ILD of the estimated BSM signals is defined as
\begin{equation}\label{ILD:BSM}
    \text{ILD}_{est}(f_0,\Omega_l) = 10 \log_{10}\frac{\int_{f_1}^{f_2} G(f_0,f)|z^l(\Omega_l,f)|^2 df}{\int_{f_1}^{f_2} G(f_0,f)|z^r(\Omega_l,f)|^2 df}.
\end{equation}
Here, $G(f_0,f)$ represents the Gammatone filter centered at frequency $f_0$. The ILD for both the target and the estimated signals is evaluated for a single far-field sound source arriving from direction $\Omega_l$, representing a set of $L$ uniformly distributed directions on the horizontal plane $\{\Omega_l = (\theta_l=90^\circ, \phi_l)\}_{l=1}^{L}$.

\begin{figure}
 \centerline{\framebox{
 \includegraphics[width=7.8cm]{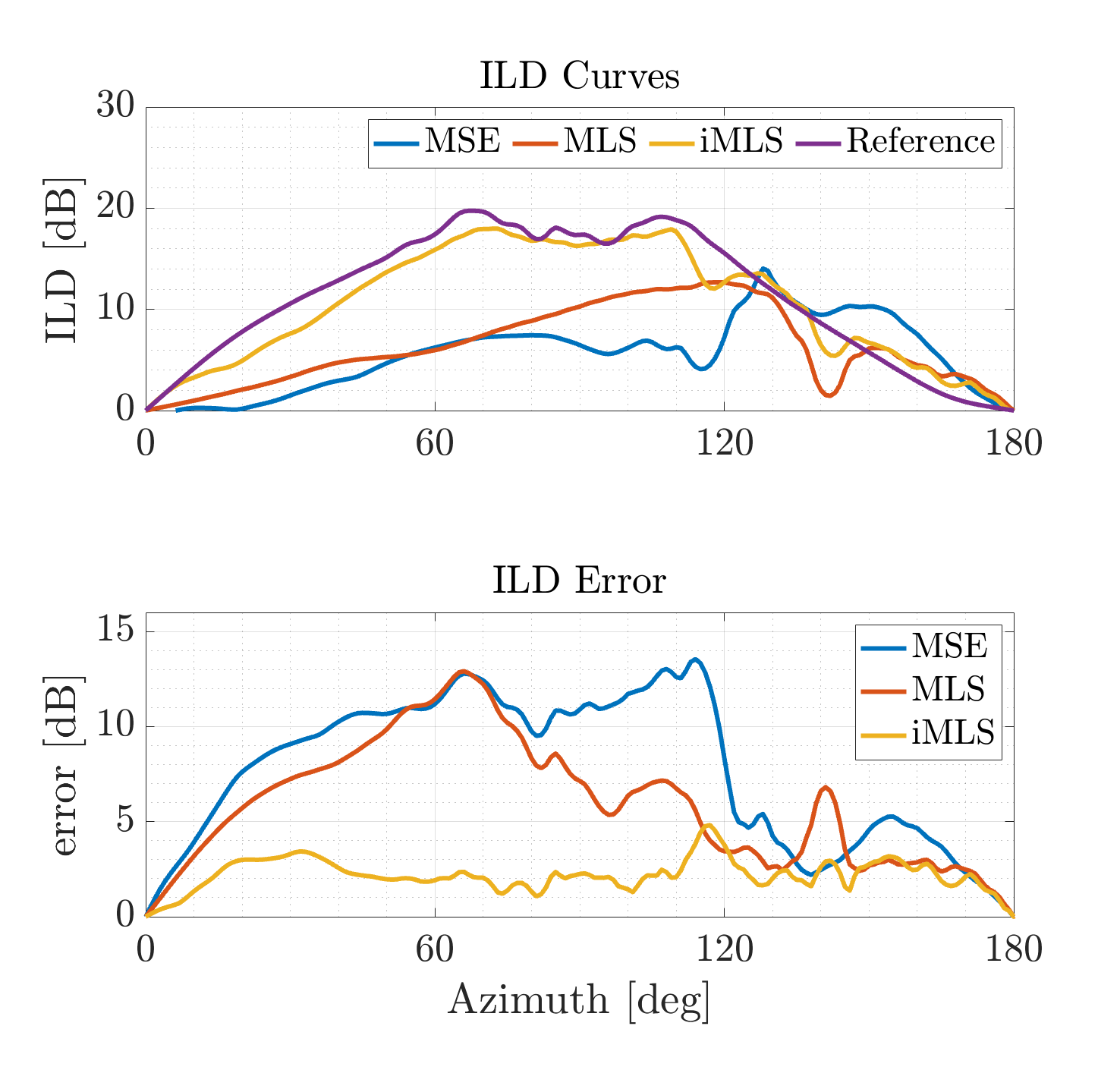}}}
 \caption{ILD curves (top) and ILD error (bottom) as a function of incident angle, averaged over the frequencies.}
 \label{fig:eILD_ang}
\end{figure}

\section{Numerical Evaluation}
This section presents an objective performance study evaluating the proposed iMagLS-BSM method and comparing it with MSE-BSM and MagLS-BSM.

For the evaluation, the recorded KEMAR~\cite{burkhard1975anthropometric} HRTF with $1,625$ directions, corresponding approximately to a Lebedev grid of order 35, was employed. A semi-circular $6$-microphone array mounted on a rigid sphere with a $10$cm radius was used, positioned at: $(\theta,\phi) =  [ (90^\circ,\pm 22^\circ),  \allowbreak  (90^\circ,\pm 45^\circ),  \allowbreak  (90^\circ,\pm 65^\circ) ]$, where $\theta$ measures from the positive vertical axis, and $\phi$ measures the angle counterclockwise from the sagittal axis. This configuration mimics wearable devices like microphones mounted on glasses or a headset. The steering matrix of the array was calculated for the same directions as the HRTF~\cite{rafaely2015fundamentals}.

Three BSM methods where evaluated:
\begin{itemize}
\item \textbf{MSE-BSM}: Coefficients $\mathbf{c}^{l,r}_{mse}(f)$ were directly calculated using the solution in Eq.~\ref{solution:mse}.
\item \textbf{MagLS-BSM}: Coefficients $\mathbf{c}^{l,r}_{mls}(f)$ from Eq.~\ref{problem:mls} were calculated using the variable exchange method~\cite{kassakian2006convex} over $1.5 \le f < 20$ kHz, with an initial phase of $\frac{\pi}{2}$, a tolerance of $10^{-20}$, and a maximum of $10^5$ iterations. Additionally, as suggested in~\cite{zotter2019ambisonics}, the diffuse-field covariance constraint is applied to $\mathbf{c}^{l,r}_{mls}(f)$.
\item \textbf{iMagLS-BSM}: Coefficients $\mathbf{C}_{imls}$ from Eq.~\ref{C_imls} were calculated using the Broyden Fletcher Goldfarb Shannon Quasi-Newton algorithm~\cite{broyden1970convergence} over $1.5 \le f < 20$ kHz. Note that $\epsilon^{l,r}_{mls}(f,\Omega_k)$ and $\epsilon_{ILD}(f_0,\Omega_l)$ in Eq.~\ref{C_imls} are both averaged over frequencies and directions. This choice minimizes average error over all frequencies and directions, leading to smoother error curves, potentially more perceptually favorable in terms of spectral artifacts and noise.
\end{itemize}

The estimated binaural signals ($z_{mse}^{l,r},z_{mls}^{l,r},z_{imls}^{l,r}$) were calculated with Eq.~\ref{est_signals} for free-field single sound source ($Q=1$) sound-fields, for each of the $1,625$ HRTF directions. Target binaural signals $p^{l,r}(f)$ were evaluated similarly using Eq.~\ref{traget_signals}. Note that in this evaluation, error is evaluated only in the $1.5 \le f \le 20$ kHz range since, in practice, binaural signals should be rendered using the MSE-optimized filters for the lower frequencies~\cite{zotter2019ambisonics}.

Figure~\ref{fig:eMAG_eILD_freq} (top) illustrates the normalized mean-square error, calculated for $z_{mse}^{l,r},z_{mls}^{l,r},z_{imls}^{l,r}$ using Eq.~\ref{eq:emag}. The error, presented in dB and for the frequency range $1.5 \le f \le 20$ kHz, is normalized by the magnitude of the target signal and averaged over all $1,625$ $\Omega_k$ directions, and ears. The blue, red, and yellow lines represent the magnitude error of $z_{mse}^{l,r},z_{mls}^{l,r},z_{imls}^{l,r}$, respectively. Results show that, for this frequency range, all the signals exhibit high complex error, which is in line with findings of previous studies~\cite{madmoni2020beamforming}.

Figure~\ref{fig:eMAG_eILD_freq} (middle) illustrates the normalized magnitude error, calculated for $z_{mls}^{l,r},z_{imls}^{l,r}$ using Eq.~\ref{eq:emag}. The error curves are presented in a similar manner to that of the top figure.  Results show that, as expected, $z_{mls}^{l,r}$ demonstrates the lowest magnitude error; the proposed $z_{imls}^{l,r}$ exhibits only a slightly higher error, within the range $1.5\le f \le 5$ kHz, where the magnitude error remains below $-10$ dB, and can be considered low. Performance of $z_{mse}^{l,r}$ is the worst, as expected.

Figure~\ref{fig:eMAG_eILD_freq} (bottom) depicts the ILD error of Eq.~\ref{eq:eILD} as a function of frequency, averaged over $L=360$ directions $\Omega_l$ on the horizontal plane. Note that only angles for $0^\circ-180^\circ$ are presented due to the symmetry of the sagittal plane. A notable improvement is observed in the ILD error for the proposed method, with an average decrease of $3.8$ dB throughout the frequency range compared to the MagLS solution. Moreover, the range of $1.5-5$ kHz approaches the Just Noticeable Differences (JND) value of $\sim1$ dB\cite{yost1988discrimination}. The dependence of the ILD error on the incident angle is depicted in Fig.~\ref{fig:eILD_ang}, where we note that the improvement of the proposed method is most notable in the frontal range ($0^\circ \le \phi \le 100$). The proposed solution does not improve the ILD error at incident directions arriving from back directions, potentially due to the lack of microphones in this location.

\section{Conclusions}
This paper introduced a novel optimization method, iMagLS-BSM, aimed at overcoming the limitations of previous binaural signal matching (BSM) approaches, particularly focusing on improving spatial information deficiencies associated with wearable microphone arrays. The proposed method enhances BSM by integrating an interaural level difference (ILD) error term into the widely used MagLS method.

The numerical evaluation demonstrated that the iMagLS-BSM method significantly reduces ILD errors while maintaining comparable magnitude errors when compared to the MagLS method. These findings highlight the potential of the proposed method in mitigating spatial information shortcomings and improving the accuracy of binaural reproduction signals, especially within the constraints of wearable devices.

The results suggest that the selection of BSM coefficients should not only prioritize individual magnitude accuracy but also consider the interaural relationship between ears. This insight encourages a more nuanced approach to BSM design for enhanced spatial audio perception.

As a direction for future work, we recommend a more extensive study involving diverse HRTFs, array configurations and perceptual listening tests to further validate the claims and assess the real-world applicability of the iMagLS-BSM method. This would contribute to a deeper understanding of its effectiveness across various scenarios and user preferences.

\ifCLASSOPTIONcaptionsoff
  \newpage
\fi

\bibliographystyle{IEEEtran.bst}
\bibliography{refs.bib}

\end{document}